\documentclass[pre,twocolumn,preprintnumbers,amsmath,amssymb,showpacs,nofootinbib,floatfix]{revtex4}

\usepackage{graphicx,bm}
\usepackage[normalem]{ulem}

\makeatletter
\def\graphicscale{\twocolumn@sw{0.3}{0.4}}
\def\graphicthreescale{\twocolumn@sw{0.3}{0.4}}

\begin{document}

\title{Universal scaling effects of  a temperature gradient 
at first-order transitions}

\author{Claudio Bonati, Massimo D'Elia, Ettore Vicari}

\affiliation{Dip. di Fisica dell'Universit\`a di Pisa and INFN, 
Largo Pontecorvo 3, I-56127 Pisa, Italy}

\date{\today}

\begin{abstract}

We study the effects of smooth inhomogeneities at first-order
transitions. We show that a temperature gradient at a thermally-driven
first-order transition gives rise to nontrivial universal scaling
behaviors with respect to the length scale $l_t$ of the variation of
the local temperature $T_x$.  We propose a scaling ansatz to describe
the crossover region at the surface where $T_x=T_c$, where the typical
discontinuities of a first-order transition are smoothed out.

The predictions of this scaling theory are checked, and get strongly
supported, by numerical results for the 2D Potts models, for a
sufficiently large number of states in order to have first-order
transitions.  Comparing with analogous results at the 2D Ising
transition, we note that the scaling behaviors induced by a smooth
inhomogeneity appear quite similar in first-order and continuous
transitions.

\end{abstract}


\maketitle

%64.85.-d       Ultracold gases trapped gases
%64.60.fd       General theory of critical region behavior
%05.10.Cc       Renormalization group methods 
%05.70.Jk       Critical phenomena      
%05.70.Fh       Phase transitions: general studies 
%05.50.+q       Lattice theory and statistics (Ising, Potts, etc.)
%64.60.De 	Statistical mechanics of model systems 
%        (Ising model, Potts model, field-theory models, Monte Carlo techniques, etc.)

% ========================= BODY =========================
%\narrowtext

\section{Introduction}

The theory of phase transitions~\cite{Landau-book,Wilson-works}
generally applies to homogenous systems.  However, homogeneity is
often an ideal limit of experimental conditions.  Inhomogenous
conditions may significantly affect the experimental data at phase
transitions, requiring an understanding of their effects for a correct
interpretation.  In particular, they generally smooth out the
singularities of the thermodynamic quantities at phase transitions.
Notable examples are temperature gradient effects in general setups
(see e.g. \cite{TG1, TG2, TG3, TG4}), the gravity effects in
experimental studies of fluids~\cite{MSGH-79} (in particular at the
superfluid transition in $^4$He systems~\cite{Lipa-etal-96}) and in
the phenomenology of dense astrophysical objects \cite{star, star2},
the external confining forces in cold atom experiments~\cite{BDZ-08},
and the intrinsic space-time inhomogeneity of the quark-gluon plasma
formation in heavy-ion collisions~\cite{qgp-ref1, qgp-ref2}. However,
inhomogeneity effects should not only be considered as a drawback of
the experimental setup, but they may also give rise to interesting
peculiar phenomena at phase transitions.

In the presence of smooth inhomogeneities, we may simultaneously
observe different phases at different space regions; for example
experiments of cold atoms in optical lattices show the simultaneous
presence of Mott incompressible and superfluid phases in different
regions of the inhomogenous harmonic trap~\cite{BDZ-08}. In
noncritical regimes, away from phase transitions when correlations do
not develop long length scales, inhomogeneity effects may be
effectively taken into account by local-equilibrium approximations
(LEA), assuming a local equilibrium analogous to that of the
homogenous system at the same thermodynamic parameters.  An example is
the local-density approximation widely used to study particle systems
with an effective space-dependent chemical potential~\cite{BDZ-08,
  GPS}.  At continuous (classical or quantum) transitions, where
correlations develop large length scales, LEA cannot provide a
satisfactory description: critical modes get significantly distorted
by the inhomogeneities, which give generally rise to a further length
scale $\ell$. However, for sufficiently smooth inhomogeneities we may
still observe a peculiar universal scaling with respect to their
length scale $\ell$, controlled by the universality class of the
transition of the homogenous system~\cite{CV-09, inoscal2, inoscal3}.

In this paper we study the effects of smooth inhomogeneities at
first-order transitions, for which little is known.  First-order
transitions do not develop diverging length scales in the
thermodynamic limit, thus one may naively expect trivial behaviors
under smooth inhomogenous conditions, describable by LEA.  Instead, as
we shall see, a nontrivial scaling behavior arises even at first-order
transitions, quite similar to that expected at continuous transitions,
and controlled by universal {\em critical exponents}.  In particular,
a temperature gradient at a thermal first-order transition gives rise
to a nontrivial scaling behavior with respect to its length scale $l_t
\sim T_x|\nabla_x T_x|^{-1}$. We study the universal aspects of this
scaling, which manifests themselves near $T_x\approx T_c$ when $l_t$
is large (in the infinite volume limit), the typical singularities of
first order transitions emerging as $l_t\to\infty$.

The paper is organized as follows.  In Sec.~\ref{model} we present the
model that we consider as theoretical laboratory where we develop our
scaling theory, i.e. the 2D Potts model in the presence of a
temperature gradient. 
In Sec.~\ref{seclea} we look at the behavior around the region
corresponding to the first-order transition, and compare it with the
corresponding LEA, showing how the latter fails to describe the
crossover between the two phases. In Sec.~\ref{scaans} we put forward
scaling ansatzes to describe this behavior across first-order
transitions, which extend analogous scaling phenomena expected at
continuous transitions. In Sec.~\ref{scanum} we present a numerical
analysis of the 2D Potts model which supports our scaling theory.
Finally, in Sec.~\ref{conclusions} we draw our conclusions.  In
App.~\ref{ising} we report an analogous numerical analysis for the
continuous transition of the 2D Ising model in the presence of a
temperature gradient.

\section{The model}
\label{model}

For the sake of demonstration, as simple statistical models undergoing
first-order transitions, we consider the two-dimensional (2D)
$q$-state Potts models for which several exact results are known, such
as the critical temperature, the latent heat, etc.~\cite{Wu-82}.  We
consider inhomogenous Potts models defined by the partition function
\begin{eqnarray}
Z = \sum_{ \{ s_{\bf r} \} }  e^{-H},\quad
H =  - \sum_{ i, \hat\mu} \frac{J_{{\bf r}_i}}{2} \, 
\delta(s_{{\bf r}_i}, s_{ {\bf r}_i+\hat\mu }), 
\label{pottsx}
\end{eqnarray}
where ${\bf r}_i$ are the sites of a square lattice, $s_{{\bf r}_i}$
are integer variables $1\le s_{{\bf r}_i} \le q$, $\delta(a,b)=1$ if
$a=b$ and zero otherwise, and $\hat\mu$ denotes denotes the unit vectors along
the $x$ and $y$ axes with positive and negative directions.  We assume that
$J_{\bf r}$ depends on the position, thus mimicking an effective
space-dependent inverse temperature $J_{\bf r} = T_{\bf
  r}^{-1}$. In the homogenous case, i.e.  $J_{\bf r}\equiv J$, the
square-lattice Potts model undergoes a phase transition
at~\cite{Baxter-book} $J_c = T_c^{-1} = \ln(1+\sqrt{q})$, which is
continuous for $q\le 4$ and first order for $q>4$. For $q=2$ the Potts
model becomes equivalent to the Ising model.

In our study we consider anisotropic lattices with $-L_x-1 \le x \le
L_x+1$ and $1\le y \le L_y$, and assume translation invariance along
the $y$ direction, thus $J_{\bf r}\equiv J_x$.  We consider a
power-law space dependence:
\begin{equation}
T_x = J_x^{-1} = T_c \left( 1 + \frac{x|x|^{p-1}}{l_t^p} \right),
\quad [l_t>L_x,\; p\ge 1], 
\label{tx}
\end{equation}
so that $\Delta T_x\equiv T_x-T_c$ changes sign passing from the
high-$T$ disordered phase ($x>0$) to the low-$T$ ordered phase
($x<0$).  $l_t$ provides the length scale of the space variation of
$T_x$.  
Of course, the most interesting case is the linear variation,
i.e. $p=1$.  In this case no \emph{a priori} knowledge of the $T_c$
value is required. However we consider also generic $p>1$ to crosscheck the
theoretical predictions.  We choose periodic boundary conditions along
$y$, and set $s_{-L_x-1,y} = 1$ and $T_{L_x+1} = \infty$.  Note that
in the $p\to\infty$ limit with $l_t>L_x$ we recover a homogenous model
in a slab $2 L_x \times \infty$ with mixed boundaries, i.e. fixed and
disordered.

We are interested in the infinite-volume limit so that only the length
scale $l_t$ is left. Thus we choose $L_x$ and $L_y$ sufficiently large
to make finite-size effects negligible in the region of interest,
around $x=0$.

We want to understand what happens across the first-order transition,
around $x=0$. For this purpose we consider the {\em wall} energy
density and magnetization
\begin{eqnarray}
&&e(x) = - \frac{\sum_y \langle \, {\cal E}(x,y) \,\rangle}{L_y},
\quad {\cal E}(x,y) = \delta(s_{x,y},s_{x,y+1}),
\label{emx}\\
&&m(x) = \frac{\sum_y \langle \; {\cal M}_1(x,y) \; \rangle}{L_y}  ,
\quad 
{\cal M}_n({\bf x}) = 
{q \delta(s_{\bf x},n) - 1\over q-1}. \nonumber
\end{eqnarray}
In homogenous systems $e(x)$ and $m(x)$ equal the half energy density
$E/2\equiv \langle H\rangle/(2JV)$ and the magnetization $M$,
respectively.  We also consider wall-wall connected correlations
\begin{eqnarray}
&&P_e(x_1,x_2) = \frac{1}{L_y} \sum_{y_1,y_2} \langle {\cal E}(x_1,y_1) 
{\cal E}(x_2,y_2) \rangle_c , \quad \label{defpem}\\
&&P_m(x_1,x_2) = \frac{1}{L_y} \sum_{y_1,y_2} \langle {\cal M}_1(x_1,y_1) 
{\cal M}_1(x_2,y_2) \rangle_c . \nonumber
%\\&&C_e(x) = {1\over L_y} \sum_{y_1,y_2} \langle {\cal E}(x,y_1) 
%{\cal E}(x,y_2) \rangle_c , \quad \label{defce}\\
%&&C_m(x) = {1\over L_y} \sum_{y_1,y_2} \langle {\cal M}_1(x,y_1) 
%{\cal M}_1(x,y_2) \rangle_c , \quad \label{defcm}
\end{eqnarray}

\begin{figure}[tbp]
\includegraphics*[scale=\graphicscale]{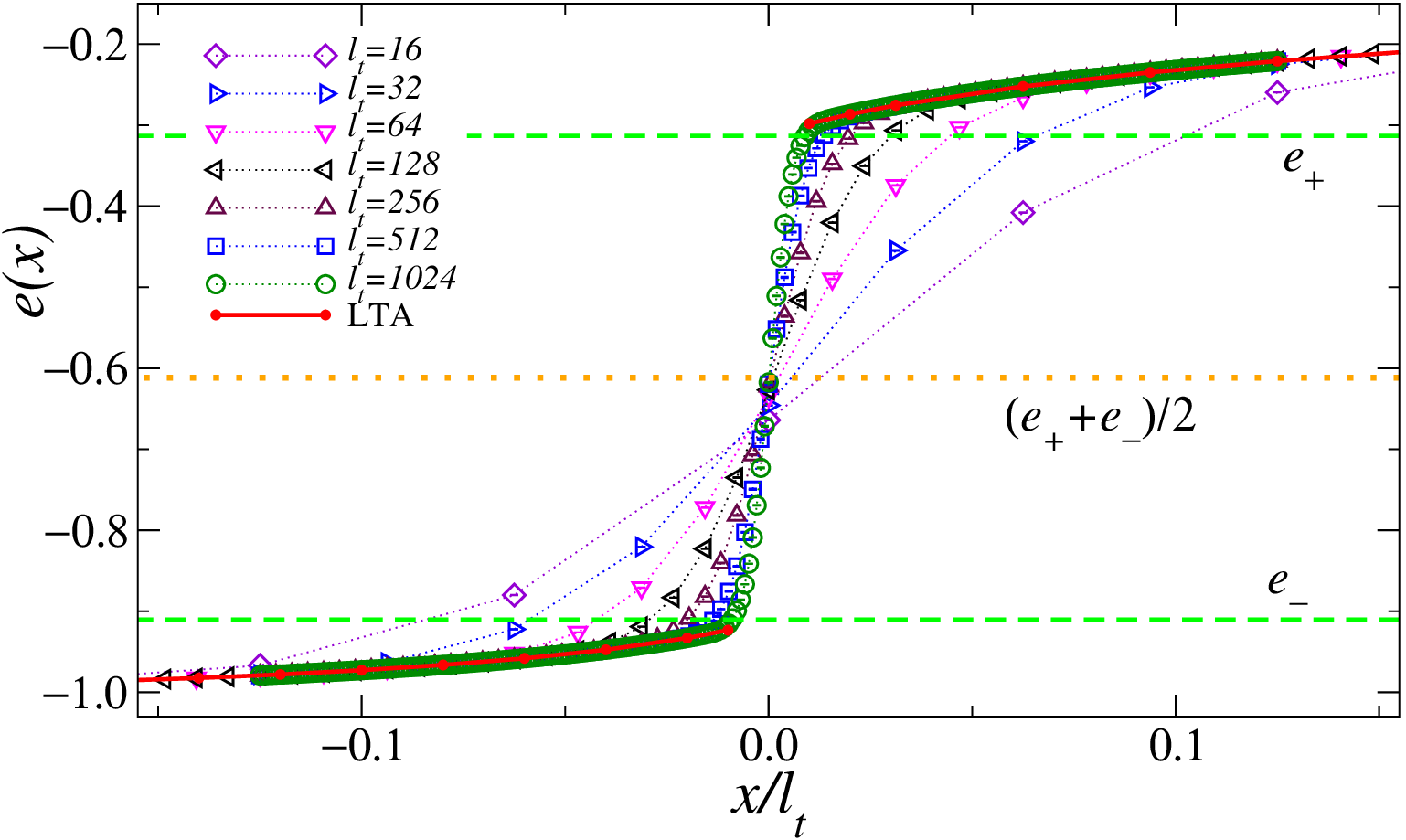}
\includegraphics*[scale=\graphicscale]{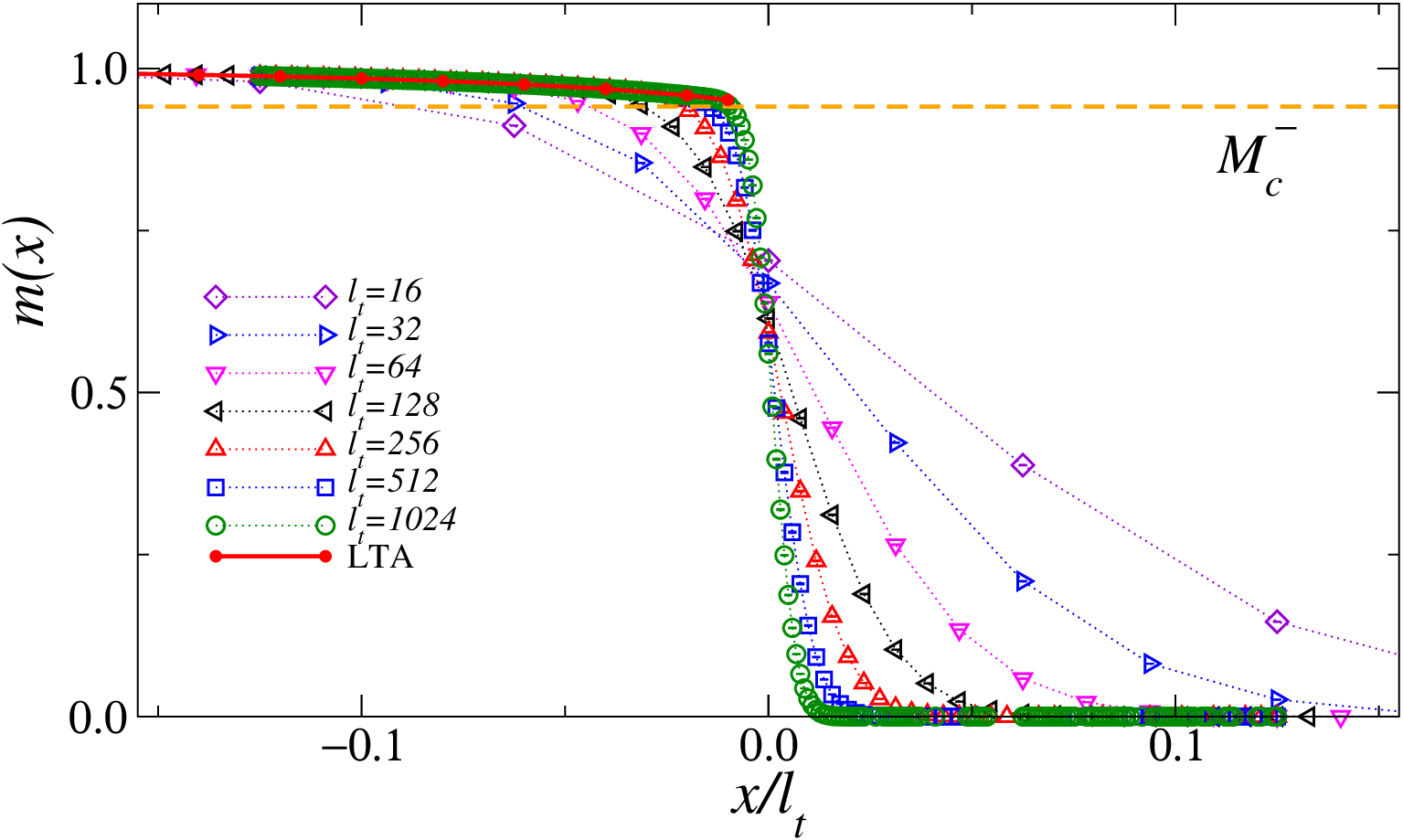}
%=% \vskip-5mm
\caption{(Color online) The wall energy density $e(x)$ and
  magnetization $m(x)$, defined in Eq.~(\ref{emx}), for $q=20$ in the
  case of a linear variation of $T_x$, i.e. $p=1$ in Eq.~(\ref{tx}),
  for several values of $l_t$.  The dashed lines represent the
  constants $e_\pm\equiv E_c^\pm/2$ (top figure) and $M_c^-$ (bottom
  figure).  The LTA data are hardly distinguishable from the
  large-$l_t$ data.  }
\label{em20}
\end{figure}

\section{Behavior across the transition region and
local temperature approximation}
\label{seclea}

We present numerical results for $q=20$, for which, beside $T_c$, we
exactly know the energy densities $E_c^-\approx-1.820684$,
$E_c^+\approx -0.626529$, and the magnetization $M_c^-\approx 0.9411759$ at
$T_c^\pm$~\cite{Wu-82}.  Monte Carlo (MC) simulations of the model
(\ref{pottsx}) are performed using a Metropolis algorithm to update
the site variables, up to length scales $l_t=O(10^3)$.

We check the convergence of the results with respect to the lattice
sizes $L_x$ and $L_y$, so that all data around $x=0$ that we present
should be considered as infinite-volume results.  In particular, for
$p=1$, we checked that $l_t/L_x = 2,\,4$ and $L_y/L_x=10,\,20$ do not
give appreciable differences (within the errors) for any observable
considered, independently of the combination of their choice. In the
case $p=2$ we mostly use $l_t/L_x=2$ and $L_y/L_x=10,\,20$.  Note that
finite-size effects around $x=0$ should be generally controlled by the
ratio $L_\#/l_t^\theta$, because $\xi\sim l_t^\theta$ in the crossover
region (the $\theta$ exponent will be defined in the following).

Figure~\ref{em20} shows MC data of $e(x)$ and $m(x)$ in the case of a
linear $T_x$.  For any $l_t$ they vary from the low-$T$ ($x<0$) to the
high-$T$ ($x>0$) regimes, showing a crossover around $x=0$.  With
increasing $l_t$, the data appear to reconstruct the discontinuities
of the first-order transition of the homogenous system.

We compare the MC results with a local-temperature approximation
(LTA), which estimates $e(x)$ and $m(x)$ using the corresponding
$E(T)/2$ and $M(T)$ of the homogenous system in the thermodynamic
limit, i.e.
\begin{eqnarray}
e(x) \approx e_{\rm lta}(x/l_t)=E[T_x(x/l_t)]/2,
\label{elta}
\end{eqnarray}
and $m(x) \approx m_{\rm lta}(x/l_t) =M[T_x(x/l_t)]$. Of course,
$e_{\rm lta}(x)$ and $m_{\rm lta}(x)$ are not defined at $x=0$,
because they are discontinuous at $T_c$.  LTA is expected to provide a
good approximation when $T_x$ varies smoothly, thus for large $l_t$.
Since $E(T)$ and $M(T)$ are not known for $T\ne T_c$, we compute them
by MC simulations of the homogenous system.  These LTA results are
also shown in Fig.~\ref{em20}.

The MC results show that $e(x)$ and $m(x)$ approach their LTA when we
consider the limit $l_t\to\infty$ keeping the ratio $x/l_t$ fixed,
i.e. at fixed $T_x$.  The convergence to LTA is fast far from $x=0$
while it becomes significantly slower when approaching $x=0$.  As we
shall see, this reflects a hidden nontrivial scaling behavior which
characterizes the crossover region around $x=0$ in the smooth
(i.e. large $l_t$) limit.  This is a novel regime probing the mixed
phase at first-order transitions, where $e_-<e(x)< e_+$ ($e_\pm \equiv
E_c^\pm/2$) and $0<m(x)<M_c^-$.

\section{Scaling ansatzes}
\label{scaans}

\begin{figure}[tbp]
\includegraphics*[scale=\graphicscale]{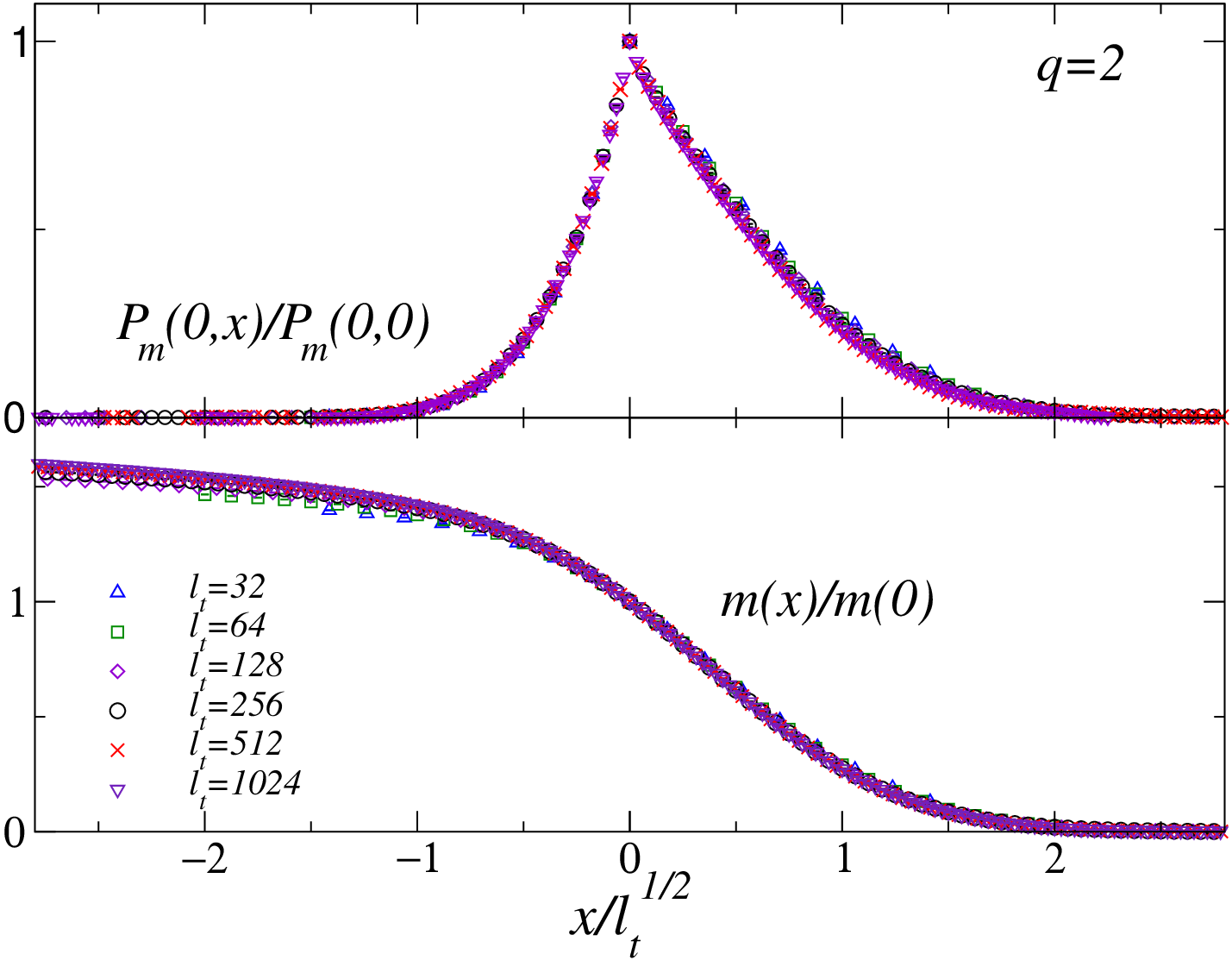}
%=% \vskip-5mm
\caption{(Color online) Scaling of $m(x)$ and $P_m(0,x)$, defined in
  Eqs.~(\ref{emx}) and (\ref{defpem}) respectively, at the continuous
  Ising transition (i.e. $q=2$) for a linear $T_x$.  The data collapse
  toward an asymptotic curve when they are plotted versus
  $x/l_t^{\theta}$ with $\theta=1/2$.  The data at $x=0$ show the
  power-law behaviors $m(0)\sim l_t^{-1/16}$ and $P_m(0,0)\sim
  l_t^{3/8}$, in agreement with the Eqs.~(\ref{mxis}) and
  (\ref{mpxis}).  }
\label{q2masca}
\end{figure}

In order to construct a theory which allows us to predict the scaling
behavior in this crossover region, we first outline the main features
of the scaling behavior at continuous transitions in analogous
conditions, such as the 2D Potts model (\ref{pottsx}) for $q\le 4$.
We assume a standard continuous transition with two relevant
parameters: $t\equiv T/T_c-1$ and $h$ 
coupled to the order parameter.  The critical behavior is
determined by their RG dimensions: for a generic $d$-dimensional
system $y_t\equiv 1/\nu$ and $y_h = (d+2-\eta)/2$, where $\nu$ and
$\eta$ are the correlation-length and two-point function critical
exponents~\cite{PV-02}. For example, $\nu=1$ and $\eta=1/4$ in the
case of the 2D Ising model. In the presence of an effective $T_x$
varying as in Eq.~(\ref{tx}), the behavior around the surface $x=0$
where $T_x=T_c$ can be inferred using scaling arguments such as those
reported in Refs.~\cite{MSGH-79,CV-09}.  Some details of their
derivation are reported in Appendix \ref{ising}.  Extending to generic
($d-1$)-dimensional walls the definitions of the observables
(\ref{emx}-\ref{defpem}), we obtain the large-$l_t$ asymptotic
behaviors
\begin{eqnarray}
&& m(x) \approx l_t^{-\theta (d-y_h)} f_m(x/l_t^\theta),\label{mxis} \\
&& e(x) \approx e_b(x/l_t) + l_t^{-\theta (d-y_t)}
  f_e(x/l_t^\theta),
\label{eneis}\\
&& P_m(x_1,x_2) \approx 
l_t^{\theta (1-2d+2y_h)} g_m(x_1/l_t^\theta,x_2/l_t^\theta),\label{mpxis}\\
&& P_e(x_1,x_2) \approx 
l_t^{\theta (1-2d +2y_t)} g_e(x_1/l_t^\theta,x_2/l_t^\theta),\label{epxis}\\
&&\theta = p/(p+y_t) \; \le 1,
\label{thetaco}
\end{eqnarray}
where we used the fact that the power-law scaling is essentially
determined by the RG dimensions of the energy density, $y_e=d-y_t$,
and the magnetization, $y_m=d-y_h$.  The exponent $\theta$ tells us
how to rescale the distances around $x=0$ to get a nontrivial scaling
behavior, thus implying that the length scale $\xi$ of the critical
modes behaves as $\xi\sim l_t^\theta$.  The term $e_b(x/l_t)$ in the
r.h.s. of Eq.~(\ref{eneis}) is a background contribution, like that
appearing in homogenous systems~\cite{CV-10,PV-02}; in Appendix
\ref{ising} we argue that it coincides with the LTA of the energy.
The approach to the asymptotic behaviors is characterized by
$O(l_t^{-\theta})$ corrections with respect to the leading terms.
Note that the above scaling behavior has some analogies with the
Finite Size Scaling (FSS) theory for homogenous systems of size
$L^d$~\cite{FBJ-73}, with three main differences: the system is
effectively infinite, the correlation length $\xi$ has a nontrivial
power-law dependence on $l_t$ (instead of $\xi\sim L$) and the spatial
inhomogeneity of the system.  Fig.~\ref{q2masca} shows some results
for $q=2$ with a linear $T_x$, i.e. $p=1$ in Eq.~(\ref{tx}). They
definitely support the scaling behaviors predicted by
Eqs.~(\ref{mxis}) and (\ref{mpxis}).

Let us now go back to first-order transitions, focussing on the
crossover region around $x=0$, see Fig.~\ref{em20}.  Our working
hypothesis is that the analogy of the inhomogenous scaling ansatz
(\ref{mxis}) and (\ref{eneis}) with the FSS of the homogenous system
may also extend to first-order transitions.  Moreover, we recall that
the FSS at a first-order transition turns out to be similar to that at
continuous transitions, being characterized by extreme RG
dimensions~\cite{NN-75,FB-82,VRSB-93} $y_t=y_h=d$ (corresponding to
$\nu=1/d$ and $\eta=2-d$). Then, it is natural to conjecture that the
crossover region around $x=0$ at first-order transitions is also
described by the scaling behavior at continuous transitions, replacing
$y_t=y_h=d$ in Eqs.~(\ref{mxis}-\ref{thetaco}).  This
leads to the scaling behaviors
\begin{eqnarray}
&e(x) \approx f_e(x/l_t^\theta),\quad
&P_e(x_i) \approx l_t^\theta g_e(x_i/l_t^\theta),
\label{ep1o} \\
&m(x) \approx f_m(x/l_t^\theta),
\quad  &P_m(x_i) \approx l_t^\theta g_m(x_i/l_t^\theta),
\label{mp1o}\\
&
\theta = p/(d + p)\; \le 1.&
\label{theta1o}
\end{eqnarray}
Unlike Eq.~(\ref{eneis}), we do not expect background terms in the
wall energy density due to the fact that its LTA does not take values
between $e_-\equiv E_c^{-}/2$ and $e_+\equiv E_c^+/2$.  Our scaling
conjecture is quite general, i.e. it should apply to any first-order
transition with the local temperature dependence (\ref{tx}), and in
particular for a linear $T_x$ for which $\theta = 1/(d+1)$. Moreover,
we expect that the asymptotic behaviors are generally approached with
$O(l_t^{-\theta})$ corrections with respect to the leading term.

An important remark is in order: the inhomogeneous scaling ansatz at
first order transitions is by no means a trivial extension of the
continuous transition case, since in the $l_t\to\infty$ limit it must
reconstruct the peculiar singularity of the first order transition,
which is not related to a diverging length scale.

\section{Scaling across the first-order transition region in 2D Potts models}
\label{scanum} 

\begin{figure}[tbp]
\includegraphics*[scale=\graphicscale]{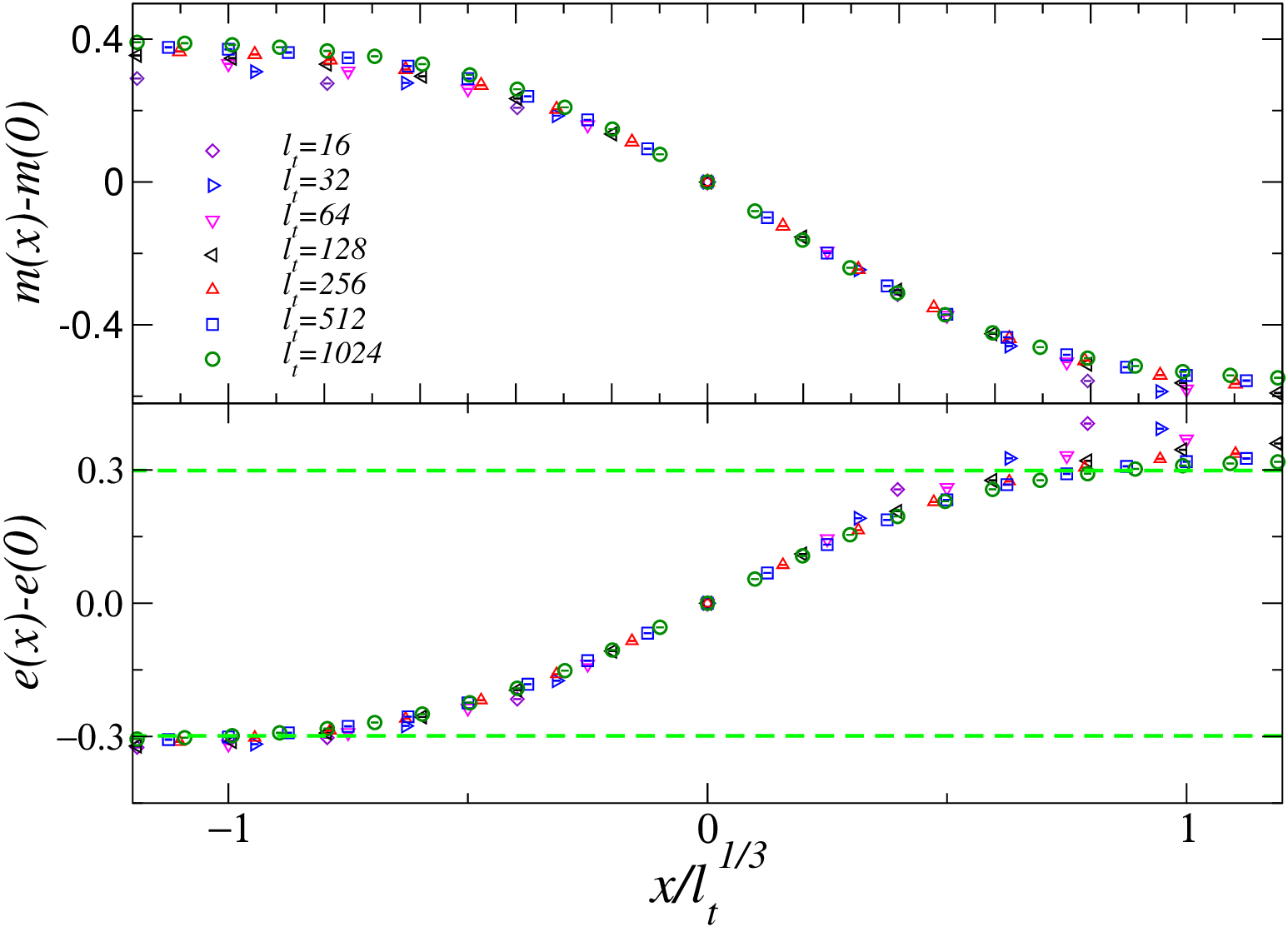}
%=% \vskip-5mm
\caption{(Color online) Scaling of the wall energy density and
  magnetization for $q=20$ with a linear $T_x$.  We
  plot the differences $e(x)-e(0)$ (bottom) and $m(x)-m(0)$ (top)
  which have smaller statistical errors.  The data clearly approach
  asymptotic curves in agreement with the scaling predictions.  In the
  bottom figure the dashed lines show the expected asymptotic values of
  the scaling curves, which are $f_e(\pm \infty)-f_e(0)= \pm
  (e_+-e_-)/2$.  }
\label{ema20r}
\end{figure}

\begin{figure}[tbp]
\includegraphics*[scale=\graphicscale]{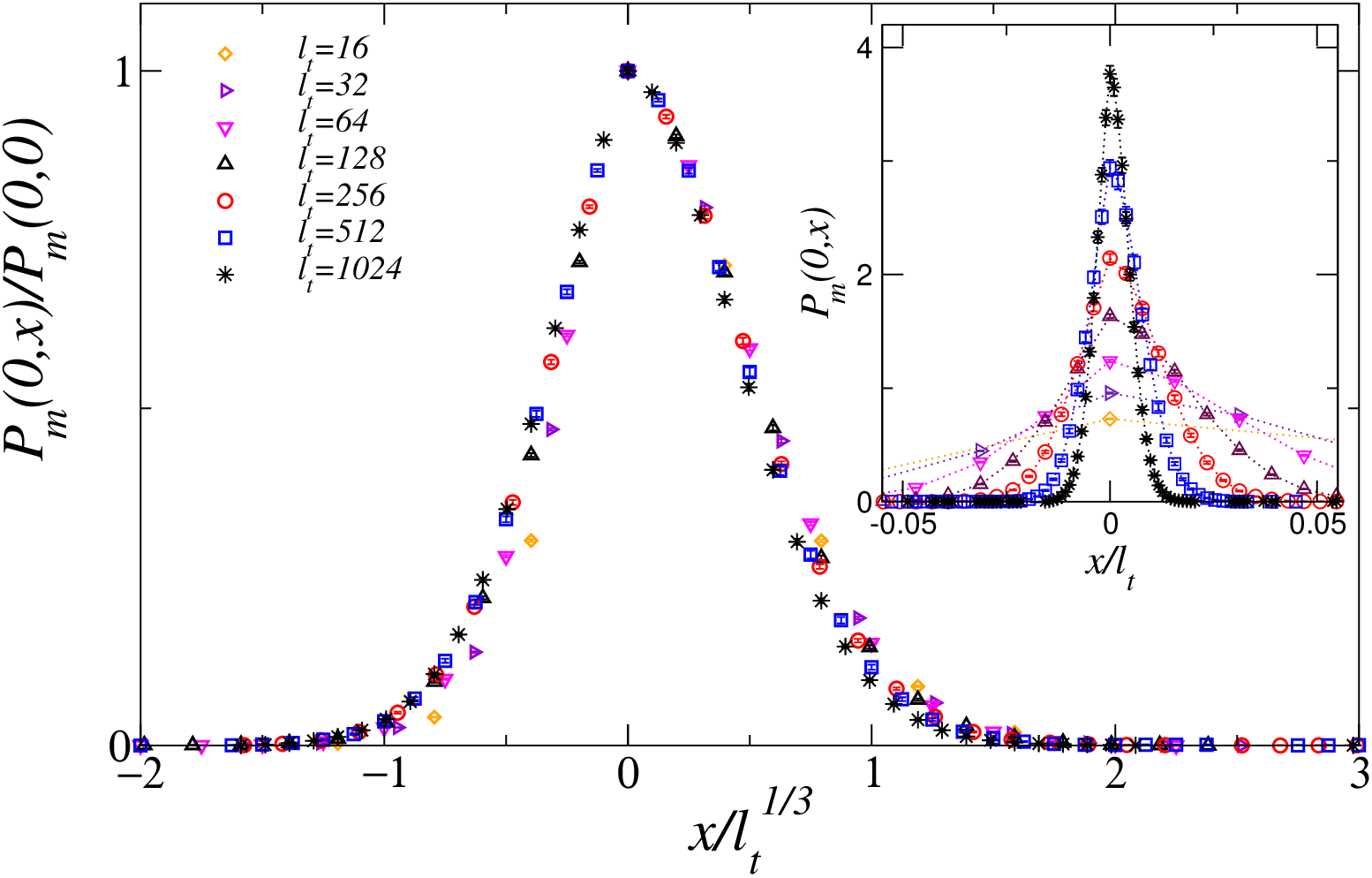}
\includegraphics*[scale=\graphicscale]{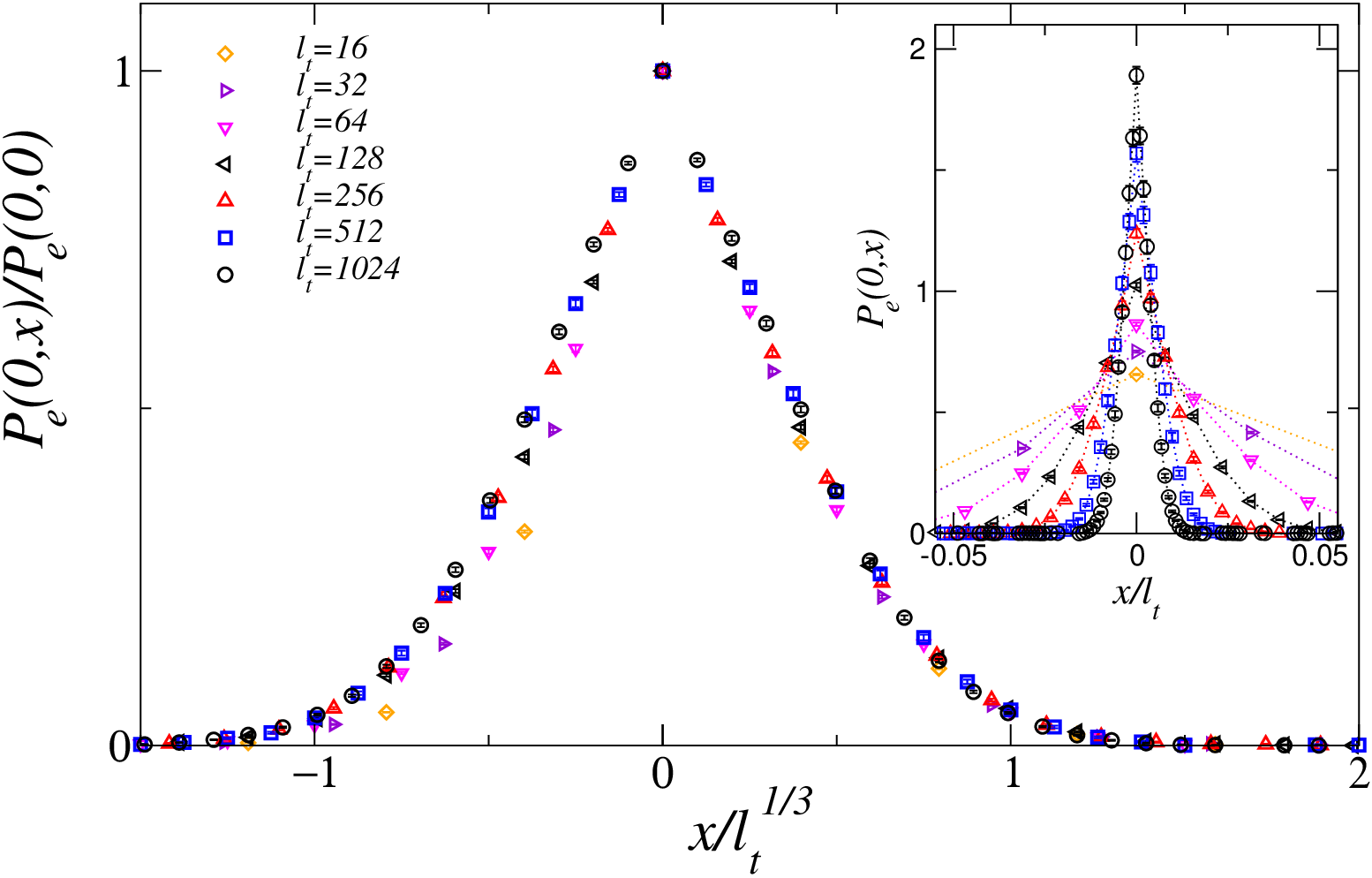}
\caption{(Color online) Scaling of the wall-wall correlations
  $P_e(0,x)$ (bottom) and $P_m(0,x)$ (top) for $q=20$ with a linear
  $T_x$.  The ratios $P_\#(0,x)/P_\#(0,0)$ versus $x/l_t^{\theta}$
  (with $\theta=1/3)$ approach scaling curves, in agreement with
  Eqs.~(\ref{ep1o}-\ref{mp1o}).  The insets show the raw data of
  $P_\#(0,x)$ vs $x/l_t$.  Note that the data also support the scaling
  predictions $P_\#(0,0)\sim l_t^{\theta}$.  }
\label{pe20}
\end{figure}

\begin{figure}[tbp]
\includegraphics*[scale=\graphicscale]{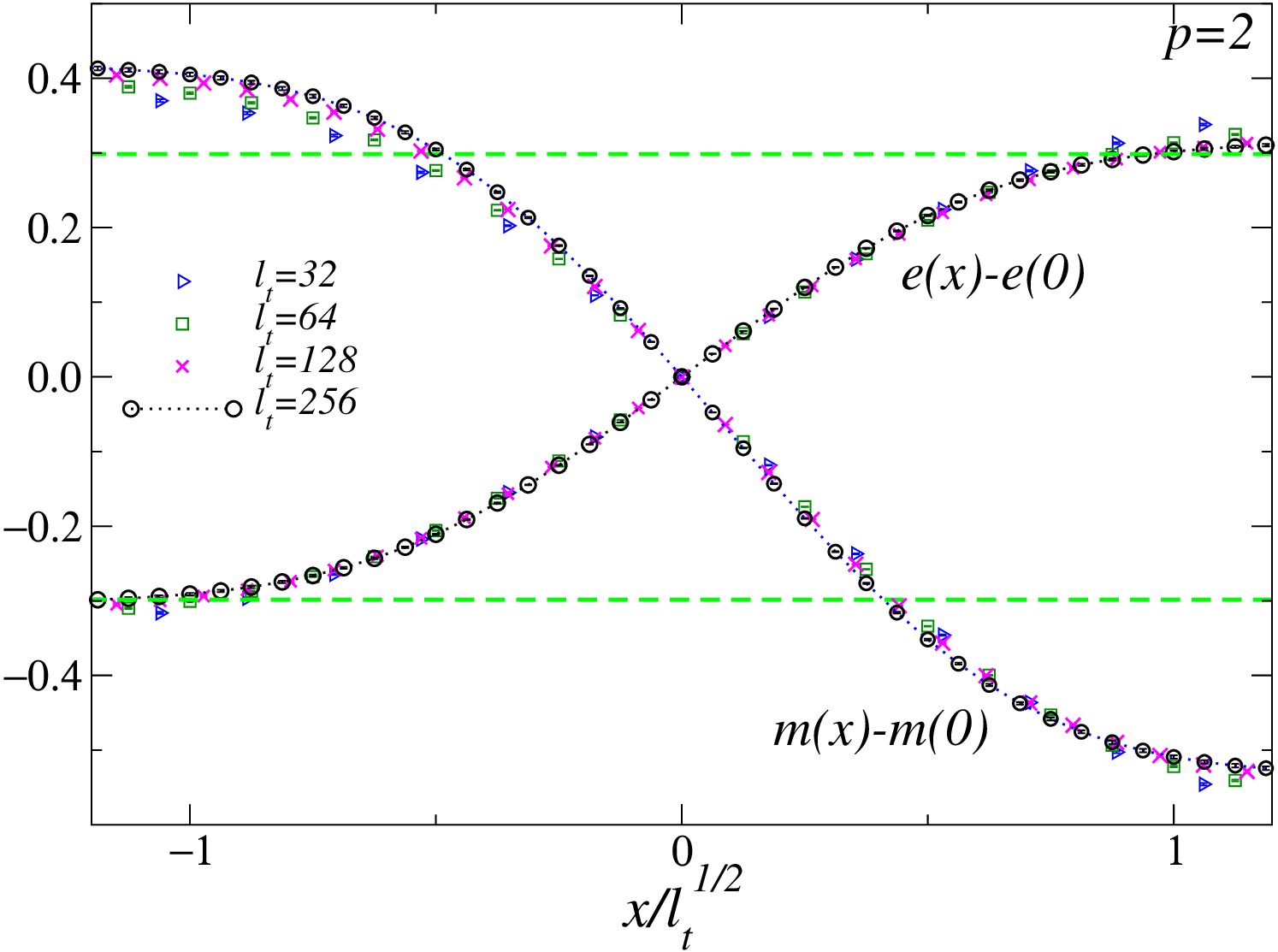}
%=% \vskip-5mm
\caption{(Color online) The wall energy density and magnetization for
  $q=20$ and a quadratic $T_x$, i.e. $p=2$ in Eq.~(\ref{tx}). The data
  clearly approach an asymptotic curve $f(x/l_t^\theta)$ with
  increasing $l_t$.  The dashed lines show the expected asymptotic
  values of the energy-density scaling curves, which are $f_e(\pm
  \infty)-f_e(0)= \pm (e_+-e_-)/2$.  }
\label{q20sq}
\end{figure}

In Figs.~\ref{ema20r}, \ref{pe20} and \ref{q20sq} we show data for
$p=1$ and $p=2$, for which $\theta=1/3$ and $\theta=1/2$ respectively.
The data of the wall energy and magnetization and the ratios
$P_e(0,x)/P_e(0,0)$ and $P_m(0,x)/P_m(0,0)$ clearly approach
asymptotic curves when they are plotted versus $x/l_t^\theta$, in
agreement with the scaling behaviors predicted by
Eqs.~(\ref{ep1o}-\ref{theta1o}).

Note that a smooth matching with the asymptotic behavior at fixed
$x/l_t$, approaching LTA, requires $f_e(\pm \infty) = e_\pm$, which is
also supported by the data, as shown by Figs.~\ref{ema20r} and
\ref{q20sq}. Analogous results are also obtained for $q=10$, whose
latent heat is substantially smaller ($\Delta\approx 0.696$): the
exponents are the same, but the scaling functions in
Eqs.~(\ref{ep1o}-\ref{mp1o}) differ quantitatively, as expected,
although they appear qualitatively similar. It is however important to
notice that we expect the weaker the transition the larger $l_t$ has
to be in order to observe the asymptotic scaling.

\begin{figure}[tbp]
\includegraphics*[scale=\graphicscale]{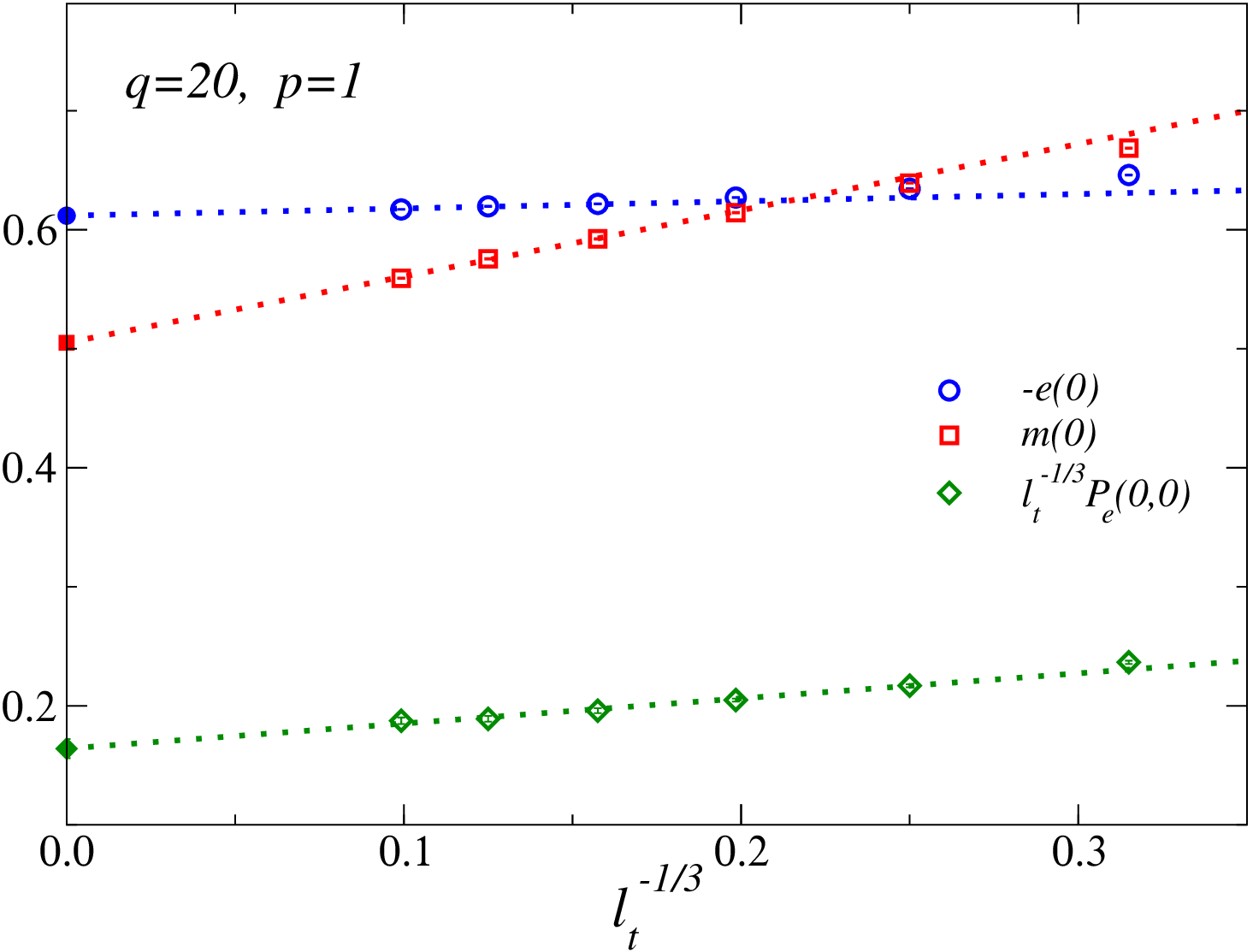}
\caption{(Color online) Large-$l_t$ behavior at $x=0$ of the wall
  energy density and magnetization and the correlation $P_e$, in the
  case of a linear variation of the temperature, whose 
  $\theta=1/3$.  The lines show fits of the data to $a+b l_t^{-1/3}$.
}
\label{em020p1}
\end{figure}

\begin{figure}[tbp]
\includegraphics*[scale=\graphicscale]{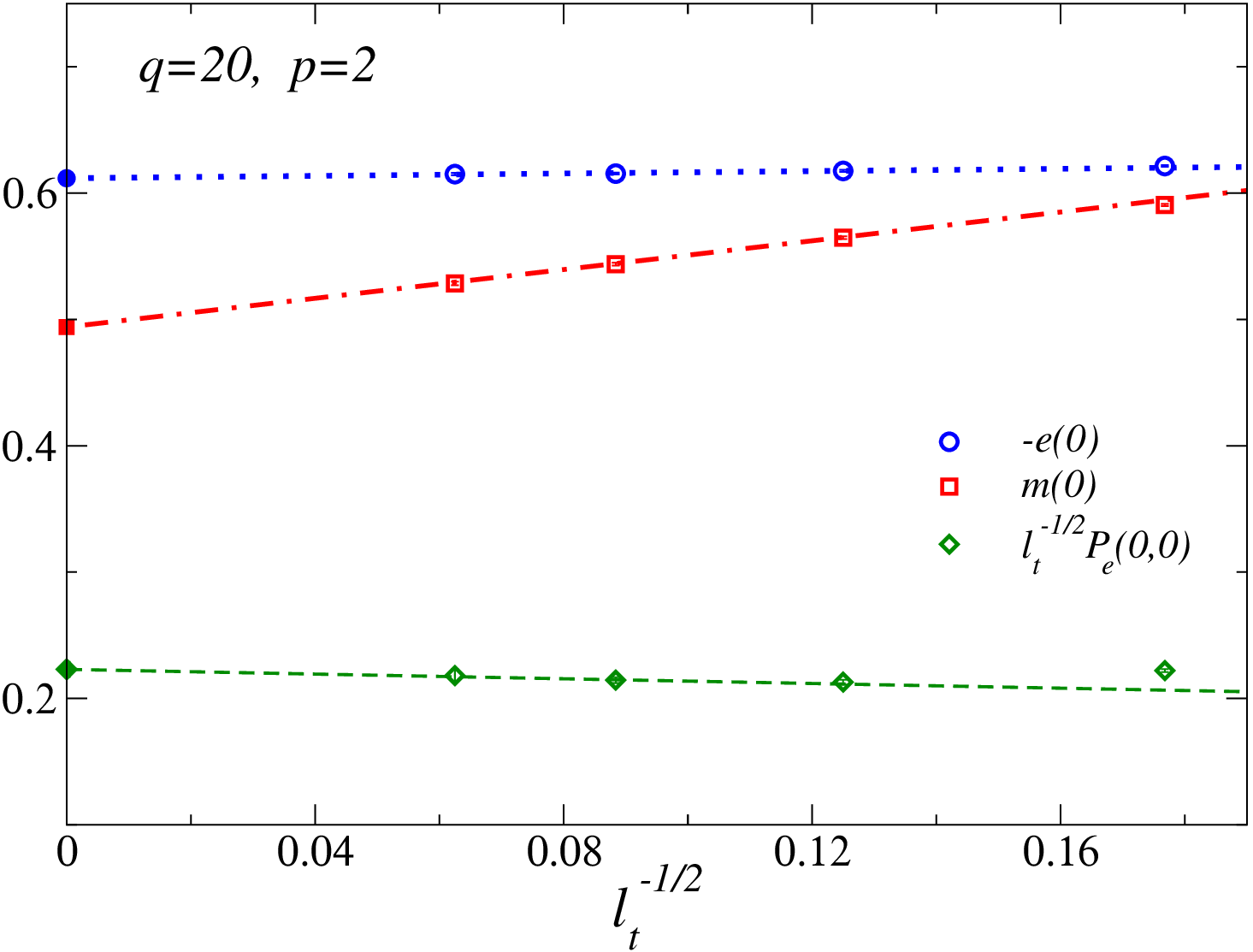}
\caption{(Color online) Large-$l_t$ behavior at $x=0$ of the wall
  energy density and magnetization and the correlation $P_e$, in the
  case of a quadratic variation of the temperature, whose 
  $\theta=1/2$.  The lines show fits of the data to $a+b l_t^{-1/2}$.
}
\label{em020p2}
\end{figure}

The approach of the data to the asymptotic behavior is also consistent
with the expected $O(l_t^{-\theta})$ corrections.  In particular, the
data of $e(x)$ support the relation $e(0) = (e_+ + e_-)/2 +
O(l_t^{-\theta})$, thus $f_e(0) = (e_+ + e_-)/2$.  A numerical
evidence of this fact can be obtained by analyzing the large-$l_t$
behavior of MC data at fixed $x/l_t^\theta$, and in particular at
$x=0$.  In Figs.~\ref{em020p1} and \ref{em020p2} we show data for the
wall energy density and magnetization and the correlation $P_e(0,x)$
at $x=0$, in the case of linear and quadratic dependence of the
temperature respectively, i.e. $p=1$ and $p=2$, corresponding to
$\theta=1/3$ and $\theta=1/2$ respectively.  The inhomogeneous scaling
behaviors (\ref{mxis}-\ref{epxis}) predict that $e(0)$ and $m(0)$ go
to a constant, while $P_e(0,0)\sim l^{\theta}$.  The data are clearly
consistent with an $O(l_t^{-\theta})$ approach to the corresponding
$l_t\to\infty$ limit. In particular, the energy density appears to
converge to the value $(e_++e_-)/2$, where $e_\pm = E_c^\pm/2$
and~\cite{Wu-82} $E_c^-\approx -1.820684$, $E_c^+\approx-0.626529$.

Note that these results imply that the curves for different values of
$l_t$ cross each other around $x=0$, as shown in Fig.~\ref{em20}, and
this crossing point approaches the line at $T_c$.  One may exploit
this property to estimate $T_c$ when it is not known, using a linear
$T_x$ (for which the line $x=0$ is not particular) and looking at the
crossing point of the wall energy density and magnetization data.  The
results are expected to approach $T_c$ with corrections corresponding
to $\Delta T \equiv T-T_c = O(l_t^{-1})$.

\section{Conclusions}
\label{conclusions}

We have shown that a temperature gradient $|\nabla_x T_x|/T_x\sim
l_t^{-1}$ induces nontrivial universal scaling behaviors at
first-order transitions driven by the temperature.  We propose a
scaling ansatz to describe, in the large $l_t$ limit, the crossover
region at $T_x\approx T_c$, where the typical discontinuities of a
first-order transition get smoothed out, and the system is effectively
probing the mixed phase.  This scaling behavior is nontrivial, and it
is such that the typical singularities of first-order transitions must
be recovered in the limit $l_t\to\infty$. We provide numerical
evidence of such phenomenon in the case of the 2D Potts models, for a
sufficiently large number of states in order to have first-order
transitions. Comparing with analogous results at the 2D Ising
transition, we note that the scaling behaviors induced by a smooth
inhomogeneity at first-order transitions appear quite similar to that
at continuous transitions.

Our approach is quite general, the results can be straightforwardly
extended to other sources of inhomogeneities smoothing out the
singularities of the transition.  For example, an analogous scaling
behavior is expected in the case the inhomogeneity arises from an
external source coupled to the order parameter (for example, this is
the case of the 2D Ising model in the low-$T$ phase with an
inhomogenous magnetic field), or when it entails a space-dependent
density in particle systems, such as in cold atom experiments. We
believe that these peculiar scaling effects of smooth inhomogeneities
at first-order transitions should be observable in experiments of
physical systems, requiring essentially the possibility of measuring
local quantities, and controlling and tuning the length-scale of the
inhomogeneity.

\bigskip

\noindent
\emph{Acknowledgement:} It is a pleasure to thank Jacopo Nespolo and 
Paolo Rossi for useful comments and discussions.

\appendix

\section{Ising model with an inhomogenous temperature}
\label{ising}

We report a detailed analysis of the scaling behavior of systems
undergoing continuous transitions in inhomogeneous conditions which
can be effectively described by a local space-dependent temperature
along one direction as in Eq.~(\ref{tx}).  This analysis can be
straightforwardly extended to variations involving more directions.

In order to compute the exponent $\theta$ associated with the
temperature inhomogeneity, we need to derive the renormalization-group
(RG) properties of the perturbation induced by the external field.
For this purpose we may follow the field-theoretical approach of
Ref.~\cite{CV-10}, and consider for simplicity the Ising universality
class which can be described by a $d$-dimensional $\Phi^4$ quantum
field theory
\begin{equation}
H_{\Phi^4} = \int d^d x                                        
\left[ \partial_\mu \phi({\bf x})^2 + 
r \phi({\bf x})^2 + u \phi({\bf x})^4\right],
\label{hphi4}
\end{equation} 
where $\phi$ is a real field associated with the order parameter, and
$r,u$ are coupling constants.  Since the temperature is related to the
energy operator $\phi^2$, we can write the perturbation $P_{T_x}$ as
\begin{eqnarray}
&&P_{T_x} = \int d^d x\, t(x) \phi({\bf x})^2,
\label{pertu}\\
&&t(x) \equiv {T_x-T_c\over T_c} \sim v^p x |x|^{p-1}.
\end{eqnarray}
Introducing the
RG dimension $y_v$ of the constant $v$, we derive the RG relation
\begin{eqnarray}
py_v - p + y_e = d,\label{rg1}
\end{eqnarray} 
where $y_e=d-1/\nu$ is the RG dimension of the energy operator. We
eventually obtain
\begin{equation}
\theta = {1\over y_v}= {p\nu\over 1 + p \nu}.
\label{thetap}
\end{equation}
This is equivalent to assuming that $t(x)$ has globally RG dimension
$y_t$, thus under a change of length scale $x\to x/b$
for which $v \to v_b$, it transforms into
\begin{equation}
b^{y_t} v^p x |x|^{p-1} = v_b^p b^{-p} x |x|^{p-1}
\label{vvb}
\end{equation}
so that $v_b = b^{1+y_t/p} v$, thus implying Eq.~(\ref{thetap}).

In particular for a linear variation, i.e. nonzero constant
temperature gradient, we have $\theta = \nu/(1+\nu)$.  Actually, this
result is very general, i.e.  it holds for any continuous transition,
replacing the appropriate exponent $\nu$.  This RG scaling analysis
leads to the following singular part of the free energy density
\begin{equation}
F = l_t^{-\theta d} {\cal F}(t l_t^{\theta y_t}, h l_t^{\theta y_h},xl_t^{-\theta}).
\label{freee}
\end{equation}
This is quite analogous to the scaling of particle systems in a
inhomogeneous trap~\cite{CV-09,CV-10}.  Note that the above scaling
behavior has some analogies with the FSS theory for homogenous systems
of size $L^d$~\cite{FBJ-73}, with two main differences: the
inhomogeneity due to the space-dependence of the external field, and
the nontrivial power-law dependence of the correlation length $\xi$
when increasing $l_t$, instead of simply $\xi\sim L$.

Then we consider generic observables defined within
translationally-invariant walls with coordinate $x$ along the
direction where the temperature varies, such as the wall energy
density and magnetization defined in Eq.~(\ref{emx}). 
Their asymptotic scaling behaviors for large
$l_t$ is expected to be
\begin{equation}
O(x) \approx l_t^{-\theta y_o} f_o(x/l_t^\theta),\label{oasy}
\end{equation}
where $y_o$ is the RG dimension of the corresponding local operator
${\cal O}$ at the fixed point describing the critical behavior of the
homogenous system.  For example, the RG dimensions of the order
parameter (magnetization) is $y_m = d - y_h$ and that of the energy
density is $y_e = d - y_t$~\cite{PV-02}.  Then for the particular
observables (\ref{emx}-\ref{defpem}) we obtain the asymptotic
behaviors (\ref{mxis}-\ref{epxis}).  Note that in some cases analytic
backgrounds arise, beside the scaling behaviors~\cite{CV-10,PV-02},
such as the case of the energy density, cf. Eq.~(\ref{eneis}).  The
approach to the asymptotic scaling behavior is characterized by
relative $O(l_t^{-\theta})$ corrections, as argued in
Ref.~\cite{CV-10} for analogous issues.

In the following we focus on the 2D Ising model, which is equivalent
to the $q=2$ Potts model (\ref{pottsx}), with a linear temperature
dependence on space, for which we have $\nu=1$, $\eta=1/4$ and
$\theta=1/2$.  An analogous study for the quantum Ising chain was
reported in Ref.~\cite{PKT-07}.  We recall that continuous transitions
occur also for $q=3,\,4$.  Their critical exponents are $\nu=5/6$ and
$\eta=4/15$ for $q=3$, $\nu=2/3$ and $\eta=1/4$ for $q=4$. Then, using
Eq.~(\ref{thetaco}) one may derive the corresponding $\theta$
exponent, e.g.  we obtain respectively $\theta=5/11$ and $\theta=2/5$
for $p=1$.

\begin{figure}[tbp]
\includegraphics*[scale=\graphicscale]{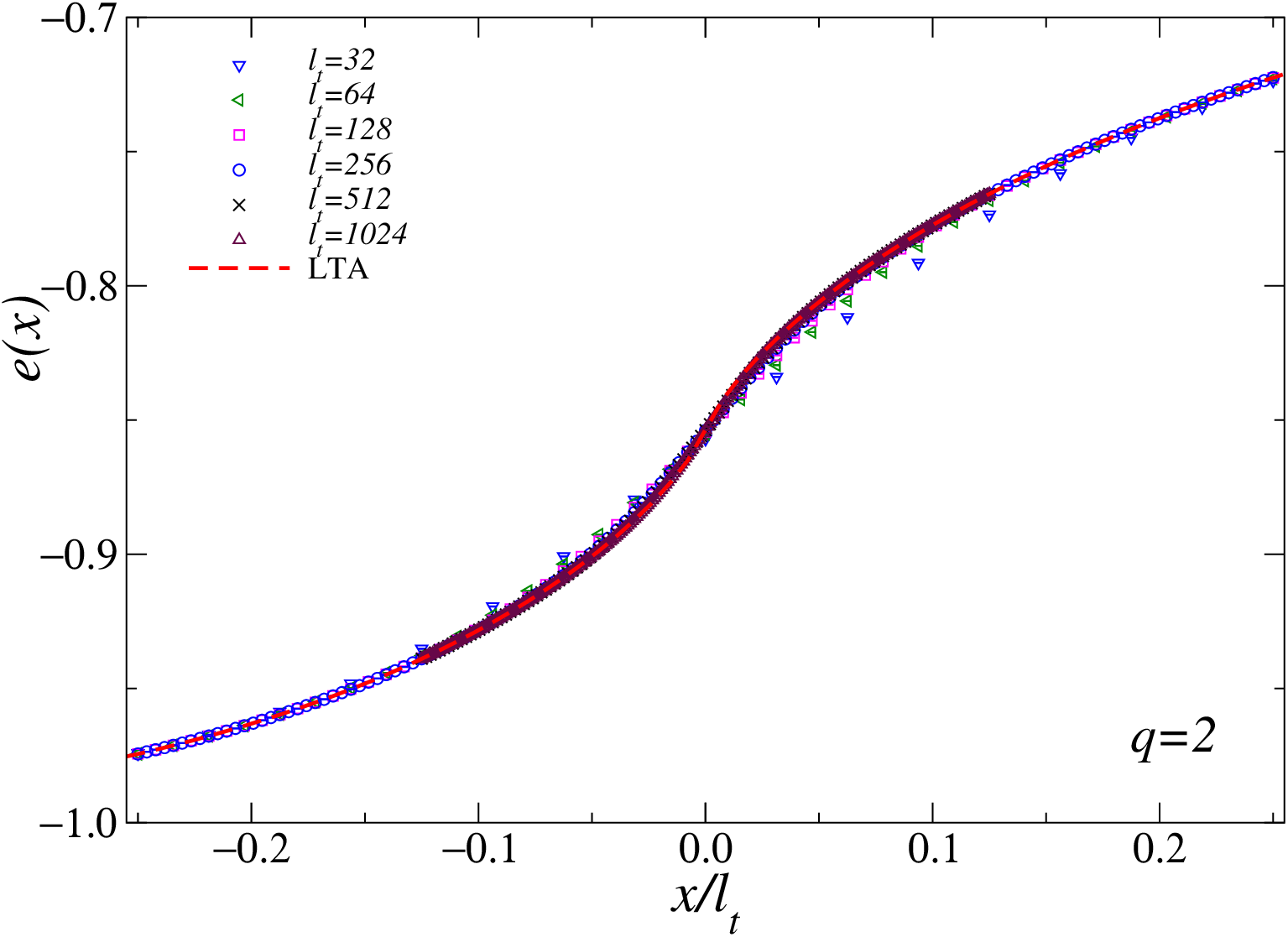}
\includegraphics*[scale=\graphicscale]{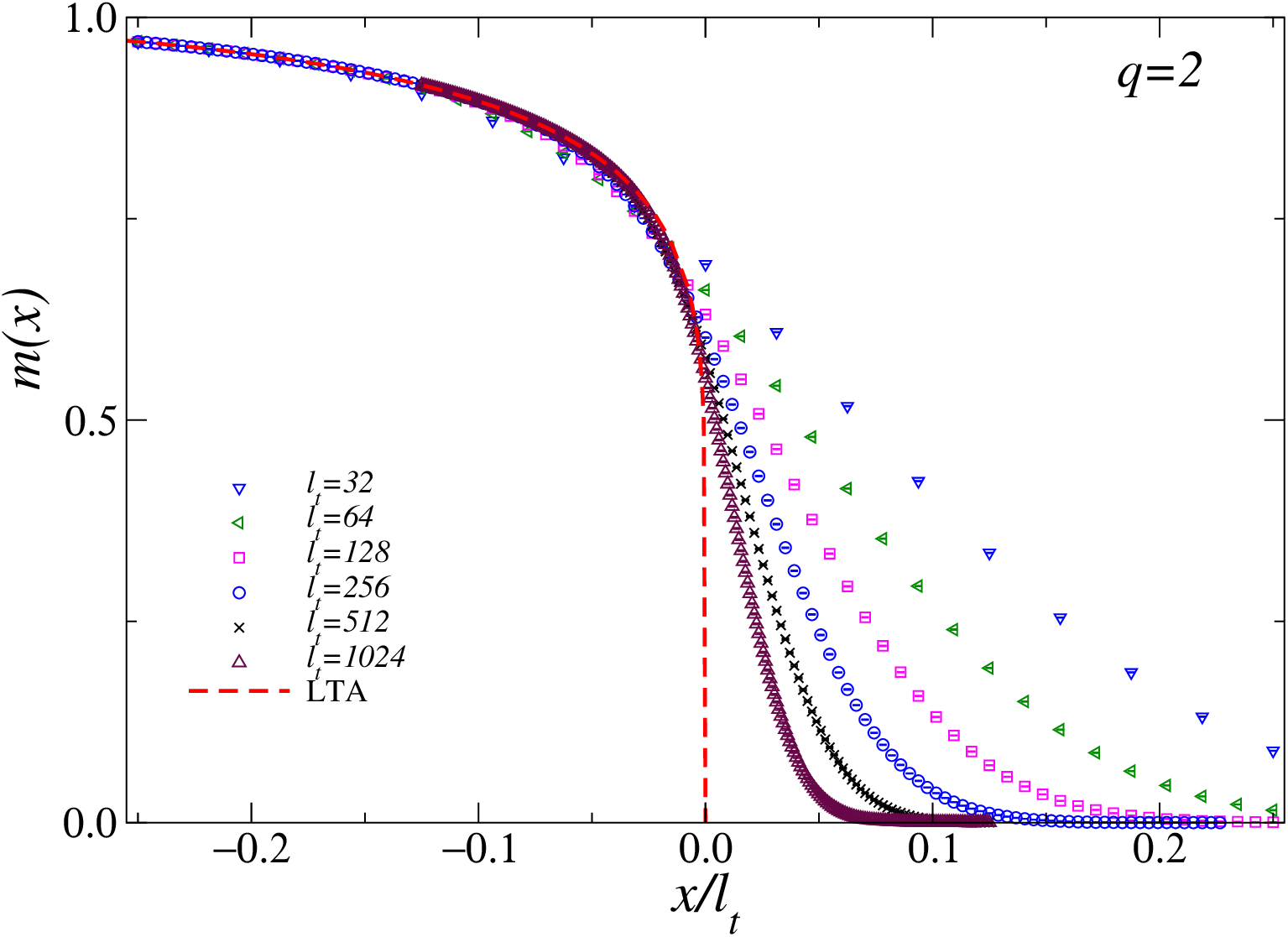}
\caption{(Color online) We show data for the wall energy density and
  magnetization for the 2D Ising model with a linear $T_x$. The dashed lines
  show the corresponding LTA,  cf. Eq.~(\ref{eltai}) and (\ref{mltai}),
  which are hardly distinguishable from the large-$l_t$ data.}
\label{emq2}
\end{figure}

\begin{figure}[tbp]
\includegraphics*[scale=\graphicscale]{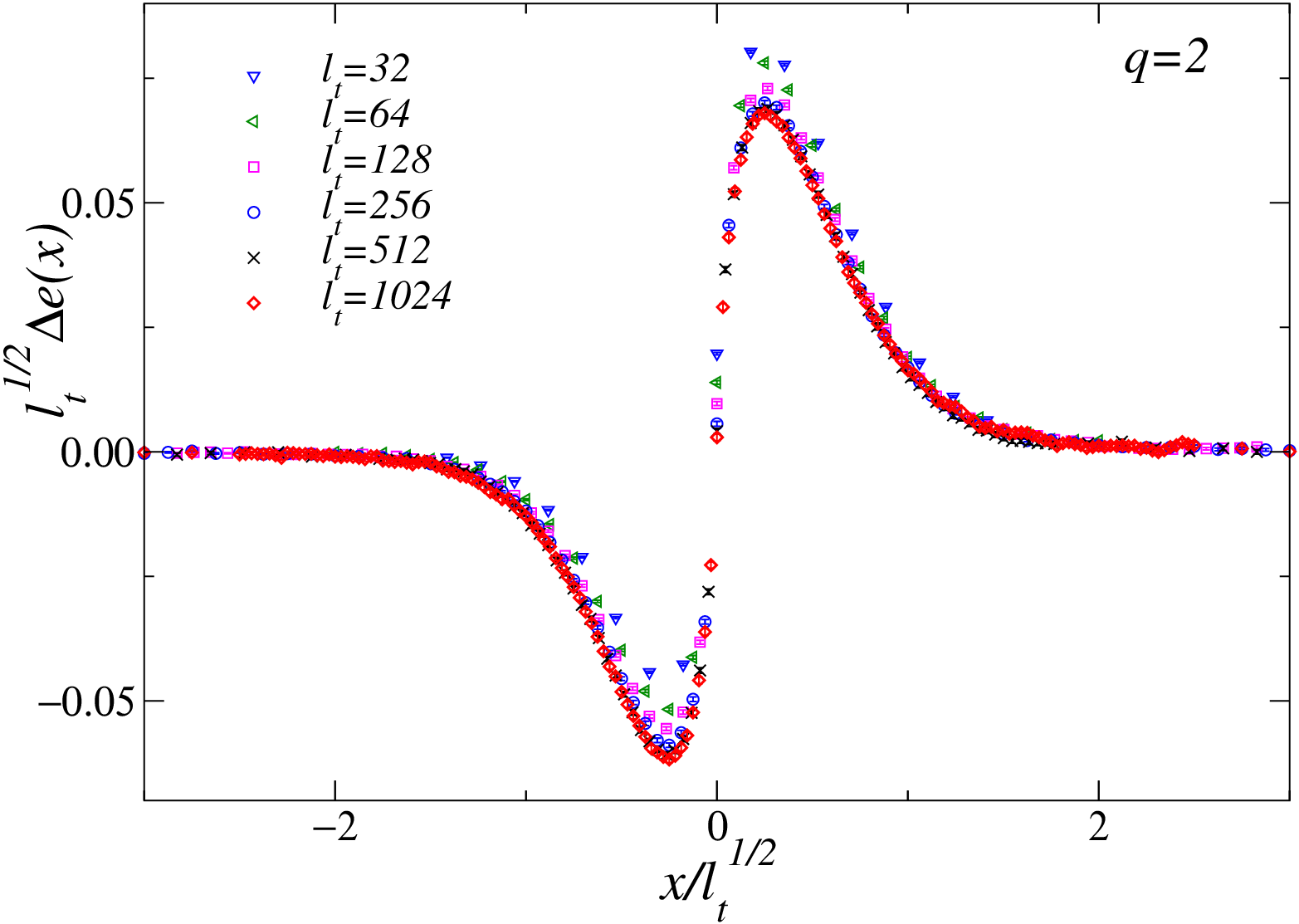}
\caption{(Color online) 
  Data for the subtracted wall energy density (\ref{deltae}).
  The data clearly converge toward an asymptotic curve $f_e(x/l_t^\theta)$,
  in agreement with the scaling behavior reported in Eq.~(\ref{eneiss}).
  The data suggest that the scaling function $f_e$ is odd.
}
\label{q2enesub}
\end{figure}

\begin{figure}[tbp]
\includegraphics*[scale=\graphicscale]{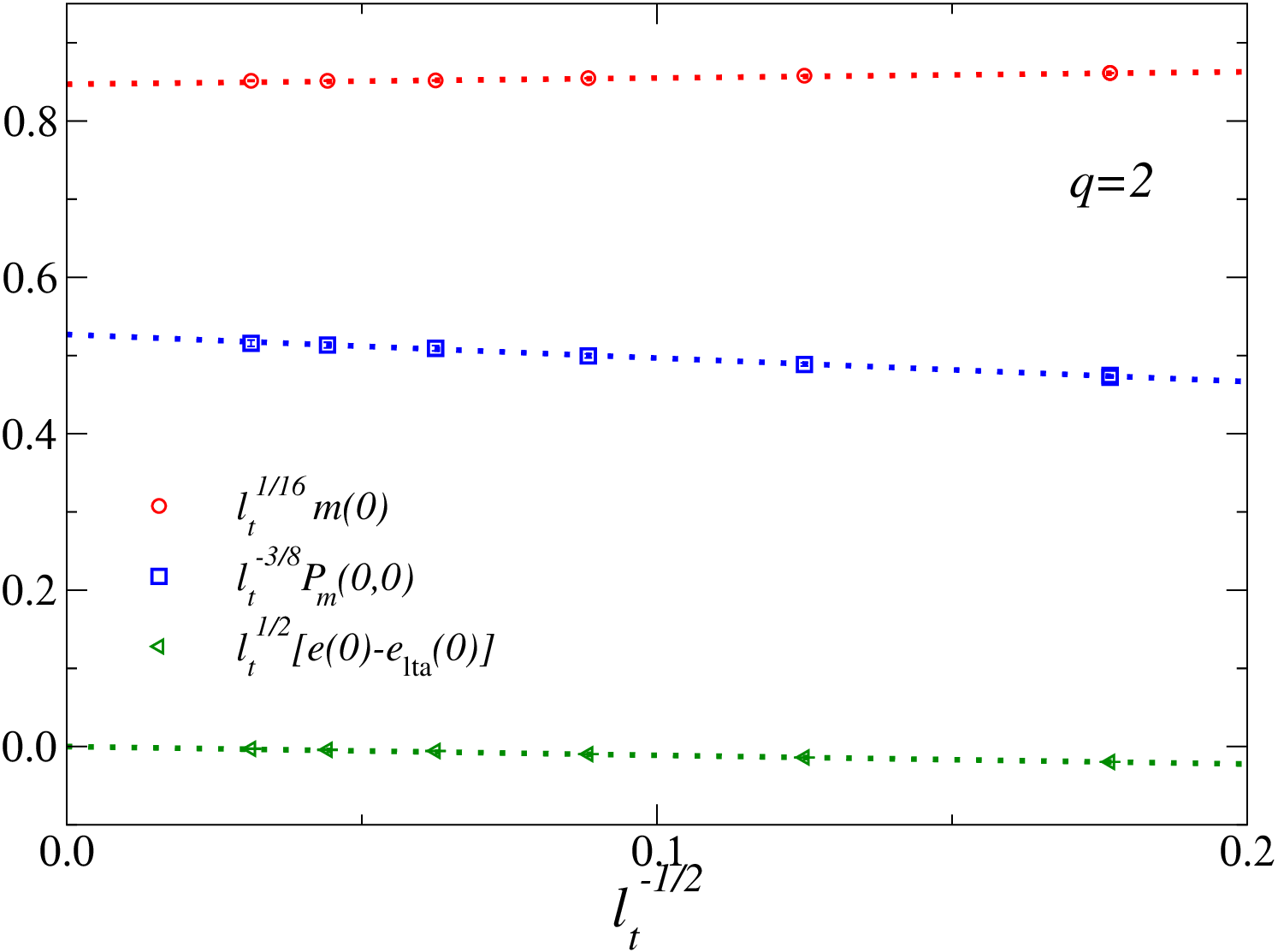}
\caption{(Color online) We check the scaling corrections of some
  quantities at $x=0$ where $T_x=T_c$.  The dotted and dashed lines show
  fits of the data to $a+b l_t^{-1/2}$.  In particular, in the case of
  the energy density the data converge to zero, thus implying
  $f_e(0)=0$, as also suggested by the numerical evidence that
  $f_e(x)$ is odd, see Fig.~\ref{q2enesub}.  }
\label{q2sca0}
\end{figure}

In Fig.~\ref{emq2} we show results for the wall energy density and
magnetization. We compare these results with the
corresponding local-temperature approximation (LTA), i.e.
\begin{eqnarray}
&&e(x) \approx e_{\rm lta}(x/l_t) \equiv E_{\rm Is}[T_x(x/l_t)]/2,
\label{eltai}\\
&&m(x) \approx m_{\rm lta}(x/l_t) \equiv M_{\rm Is}[T_x(x/l_t)],
\label{mltai}
\end{eqnarray}
where $E_{\rm Is}(T)$ and $M_{\rm Is}(T)$ are the energy density and magnetization of
the homogenous system at the temperature $T$ in the thermodynamic
limit, which are exactly known for the 2D Ising model~\cite{MW-73}. 
Setting
\begin{equation}
\beta \equiv  1/T,\qquad 
\tau\equiv  {1- \sinh(\beta)^2\over 2\sinh(\beta)},
\label{betatau}
\end{equation}
the magnetization in the low-$T$ phase reads
\begin{equation}
M_{\rm Is}(T) = \left[ 1 -   \left( \sqrt{1+\tau^2} + \tau\right)^2\right]^{1/8},
\label{magnq2}
\end{equation}
and the energy density
\begin{eqnarray}
E_{\rm Is}(T) &=& - \frac{\partial F}{\partial\beta},\label{ede}\\
F &=& \beta + \ln[\sqrt{2}\cosh(\beta)] + \label{freen}\\
&& +\int_0^\pi \frac{d\varphi}{2\pi} 
{\rm ln}\left\{ 1 + \left[ 1 - 
\frac{\cos^2\varphi}{1 + \tau^2}\right]^{1/2}\right\}.
\nonumber
\end{eqnarray}
The LTA of the energy density and magnetization,
cf. Eqs.~(\ref{eltai}) and (\ref{mltai}), are expected to improve with
increasing $l_t$. An educated guess is that LTA provides their
$l_t\to\infty$ limit keeping the ratio $x/l_t$ fixed.  This is
confirmed by the data shown in Fig.~\ref{emq2}: their convergence
appears fast far from $x=0$ coinciding with $T_c$, but becomes slower
when approaching $x=0$.  These larger deviations around $x=0$ reflect
the above-discussed scaling behavior characterized by the lenght scale
$l_t^\theta$, which sets in around $x=0$.

The data shown in Fig.~\ref{q2masca} nicely confirm the predictions
for the magnetization and its correlation. Indeed the data rapidly
approach an asympotic curve when they are plotted versus the ratio
$x/l_t^\theta$. Concerning the energy density we consider the
subtracted quantity
\begin{equation}
\Delta e(x) \equiv e(x) - e_{\rm lta}(x/l_t).
\label{deltae}
\end{equation}
We argue that
\begin{equation}
\Delta e(x) \approx l_t^{-\theta} f_e(x/l_t^\theta ),
\label{eneiss}
\end{equation}
i.e. after subtracting the corresponding LTA only the nontrivial
scaling part is left. This is shown by the MC data of the
subtracted quantity (\ref{deltae}) shown in Fig.~\ref{q2enesub}.
This allows us to write the wall energy density as
\begin{equation}
 e(x) \approx e_{\rm lta}(x/l_t) + l_t^{-\theta y_e}
  f_e(x/l_t^\theta).
\label{eneis2} 
\end{equation}

Finally, we check the approach to the asymptotic behavior.
Fig.~\ref{q2sca0} shows it for the wall energy density, magnetization
and their correlations at $x=0$, confirming the prediction that they
are $O(l_t^{-\theta})$, i.e.  $O(l_t^{-1/2})$ in this case.

\end{document}